\documentclass[12pt]{article}

\usepackage{graphicx,epsfig,epsf,amssymb}

\usepackage{scicite}

\usepackage{times}

\newenvironment{sciabstract}{%
\begin{quote} \bf}
{\end{quote}}

\newcounter{lastnote}
\newenvironment{scilastnote}{%
\setcounter{lastnote}{\value{enumiv}}%
\addtocounter{lastnote}{+1}%
\begin{list}%
{\arabic{lastnote}.}
{\setlength{\leftmargin}{.22in}}
{\setlength{\labelsep}{.5em}}}
{\end{list}}

\title{A new population of very high energy $\gamma$-ray sources
  in the Milky Way}

\author{F.~Aharonian $^{1}$, 
  A.G.~Akhperjanian $^{2}$, 
  K.-M.~Aye $^{3}$,
  A.R.~Bazer-Bachi $^{4}$,\\ 
  M.~Beilicke $^{5}$,
  W.~Benbow $^{1}$,
  D.~Berge $^{1}$,
  P.~Berghaus $^{6, \dag}$,
  K.~Bernl\"ohr $^{1,7}$,\\
  C.~Boisson $^{8}$
% ...
\\
  \\
  \normalsize{$^\ast$To whom correspondence should be addressed; E-mail: Stefan.Funk@mpi-hd.mpg.de}\\
}

\date{}

\begin{document} 




\maketitle 

{\footnotesize
\begin{enumerate}
\item {Max-Planck-Institut f\"ur Kernphysik, P.O. Box 103980, D 69029
Heidelberg, Germany}
\item{Yerevan Physics Institute, 2 Alikhanian Brothers St., 375036 Yerevan,
Armenia}
\item{University of Durham, Department of Physics, South Road, Durham DH1 3LE,
U.K.}
\item{Centre d'Etude Spatiale des Rayonnements, CNRS/UPS, 9 av. du Colonel
Roche, BP 4346, F-31029 Toulouse Cedex 4, France}
\item{Universit\"at Hamburg, Institut f\"ur Experimentalphysik, Luruper Chaussee
149, D 22761 Hamburg, Germany}
\item{Physique Corpusculaire et Cosmologie, IN2P3/CNRS, Coll{\`e}ge de France, 11 Place
Marcelin Berthelot, F-75231 Paris Cedex 05, France}
\item{Institut f\"ur Physik, Humboldt-Universit\"at zu Berlin, Newtonstr. 15,
D 12489 Berlin, Germany}
\item{LUTH, UMR 8102 du CNRS, Observatoire de Paris, Section de Meudon, F-92195 Meudon Cedex,
France}
\item{Groupe d'Astroparticules de Montpellier, IN2P3/CNRS, Universit\'e Montpellier II, CC85,
Place Eug\`ene Bataillon, F-34095 Montpellier Cedex 5, France}
\item{Laboratoire Leprince-Ringuet, IN2P3/CNRS,
Ecole Polytechnique, F-91128 Palaiseau, France}
\item{Dublin Institute for Advanced Studies, 5 Merrion Square, Dublin 2,
Ireland}
\item{Service d'Astrophysique, DAPNIA/DSM/CEA, CE Saclay, F-91191
Gif-sur-Yvette, France}
\item{Landessternwarte, K\"onigstuhl, D 69117 Heidelberg, Germany}
\item{Laboratoire d'Astrophysique de Grenoble, INSU/CNRS, Universit\'e Joseph Fourier, BP
53, F-38041 Grenoble Cedex 9, France}
\item{Unit for Space Physics, North-West University, Potchefstroom
  2520, South Africa}
\item{Laboratoire de Physique Nucl\'eaire et de Hautes Energies, IN2P3/CNRS, Universit\'es
Paris VI \& VII, 4 Place Jussieu, F-75231 Paris Cedex 05, France}
\item{Institut f\"ur Theoretische Physik, Lehrstuhl IV: Weltraum und
Astrophysik, Ruhr-Universit\"at Bochum, D 44780 Bochum, Germany}
\item{Institute of Particle and Nuclear Physics, Charles University, V
  Holesovickach 2, 180 00 Prague 8, Czech Republic} 
\item{University of Namibia, Private Bag 13301, Windhoek, Namibia}
\item{European Associated Laboratory for Gamma-Ray Astronomy, jointly
supported by CNRS and MPG}
\end{enumerate}

\begin{itemize}
\item[$\dag$]{Universit\'e Libre de 
  Bruxelles, Facult\'e des Sciences, Campus de la Plaine, CP230, Boulevard
 du Triomphe, 1050 Bruxelles, Belgium}
\item[$\ddag$]{now at Washington Univ., Department of Physics,
 1 Brookings Dr., CB 1105, St. Louis, MO 63130, USA}
\item[$\S$]{now at Department of Physics and
Astronomy, Univ. of Sheffield, The Hicks Building,
Hounsfield Road, Sheffield S3 7RH, U.K.}

\end{itemize} 
}

\begin{sciabstract}
  Very high energy $\gamma$-rays probe the long-standing mystery of
  the origin of cosmic rays. Produced in the interactions of
  accelerated particles in astrophysical objects, they can be used to
  image cosmic particle accelerators.  A first sensitive survey of the
  inner part of the Milky Way with the High Energy Stereoscopic System
  (H.E.S.S.) reveals a population of eight previously unknown firmly
  detected sources of very high energy $\gamma$-rays. At least two
  have no known radio or X-ray counterpart and may be representative
  of a new class of `dark' nucleonic cosmic ray sources.
\end{sciabstract}

Very high energy (VHE) $\gamma$-rays with energies $E >\,10^{11}$ eV
are probes of the non-thermal universe, providing access to energies
far greater than those that can be produced in accelerators on
earth. The acceleration of electrons or nuclei in astrophysical
sources leads inevitably to the production of $\gamma$-rays, by the
decay of $\pi^{0}$s produced in hadronic interactions, inverse Compton
scattering of high energy electrons on background radiation fields or
the non-thermal bremsstrahlung of energetic electrons. Several classes
of objects in the Galaxy are suspected or known particle accelerators:
pulsars and their pulsar wind nebulae (PWN), supernova remnants
(SNRs), microquasars, and massive star forming regions. VHE
$\gamma$-ray sources that have been detected in the Galaxy include
PWNs, SNRs and objects with no identified counterpart at other
energies. Such sources without counterpart are particularly
interesting because a lack of X-ray emission could indicate that the
primary accelerated particles are nucleons rather than the high-energy
electrons.  Essentially all potential Galactic sources cluster along
the Galactic plane. Thus, a systematic survey of the Galactic plane is
the best means to investigate the properties of these source classes
and to search for yet unknown types of Galactic VHE $\gamma$-ray
emitters.

Large-scale $\gamma$-ray surveys in the TeV energy regime (1~TeV =
$10^{12}$ eV) have been performed using the Milagro water-Cherenkov
detector~\cite{MilagroAllSky} and the Tibet air-shower
array~\cite{TibetAllSky}.  These all-sky instruments have only modest
sensitivity, reaching a flux limit comparable to the flux level of the
Crab Nebula, $\sim 2 \times 10^{-11}$ cm$^{-2}$ s$^{-1}$ (for $E
>1\,$TeV), in one year of observations. Both surveys covered $\sim 2
\pi$ steradian of the northern sky and resulted only in flux upper
limits. A survey of the part of the Galactic plane accessible from the
northern hemisphere ($-2^{\circ} < l < 85^{\circ}$, where $l$ denotes
Galactic longitude) was performed by the stereoscopic High Energy
Gamma Ray Astronomy (HEGRA) array of imaging Cherenkov
telescopes~\cite{HEGRA}. No sources of VHE $\gamma$-rays were found in
this survey but upper limits between 15\% of the Crab flux for
Galactic longitudes $l>30^{\circ}$ and more than 30\% of the Crab flux
in the inner part of the Milky Way were derived~\cite{HEGRASCAN}.
Until the completion of the High Energy Stereoscopic System
(H.E.S.S.)~\cite{HESS} in early 2004, no VHE $\gamma$-ray survey of
the southern sky, or of the central region of the Galaxy had been
performed. The central part of the Galaxy contains the highest density
of potential $\gamma$-ray sources. We have conducted a survey of this
region with H.E.S.S. in 2004, at a flux sensitivity of 3\% of the Crab
flux.

H.E.S.S. is an array of telescopes exploiting the imaging Cherenkov
technique. The telescopes image the Cherenkov light emitted by
atmospheric particle cascades initiated by $\gamma$-rays or
cosmic-rays in the upper atmosphere.  Measurements of the same cascade
by multiple telescopes allow improved rejection of the cosmic-ray
background and better angular and energy resolution in comparison to
single dish telescopes. H.E.S.S. consists of four atmospheric
Cherenkov telescopes, each with 107 m$^{2}$ of mirror
area~\cite{HESSOPT} and equipped with a 960 pixel photo-multiplier
tube camera~\cite{HESSCamera}.  The four telescopes are operated in a
stereoscopic mode with a system trigger~\cite{HESSTrigger}, requiring
at least two telescopes to provide images of each
cascade. H.E.S.S. has the largest field of view ($5^{\circ}$ diameter)
of all imaging Cherenkov telescopes now in operation, which yields a
considerable advantage for surveys~\cite{HESSDesign2}.

H.E.S.S. provides unprecedented sensitivity to $\gamma$-rays above
100\,GeV, below 1\% of the flux from the Crab Nebula for long
exposures.  This sensitivity has already been demonstrated by
high-significance detections of the Galactic centre
(HESS\,J1745-290)~\cite{HESSGalCen} and the SNR
RX\,J1713.7-3946~\cite{HESS1713}.
The angular resolution for individual $\gamma$-rays is better than
0.1$^{\circ}$, allowing a source position error of 30 arc sec to be
achieved, even for relatively faint sources.

We scanned the inner part of the Galactic plane (Galactic longitude
-30$^\circ$$<$l$<$30$^\circ$ and latitude
-3$^{\circ}$$<$b$<$3$^{\circ}$) with H.E.S.S. from May to July 2004.
Observations 28~minute in duration were made in steps of $0.7^{\circ}$
in longitude and steps of $1^{\circ}$ in latitude for a total
observation time of 112 hours. An average 5-standard-deviation
detection sensitivity of
$3\,\times\,10^{-11}$\,photons\,cm$^{-2}$\,s$^{-1}$ above 100 GeV
($\approx $3\% of the Crab Flux) was reached for points on the
Galactic plane. Seven promising candidate sources from the survey
were re-observed from July to September for typically 3.5 hours per
source.

\begin{figure}
\begin{center}
\includegraphics[width=\textwidth]{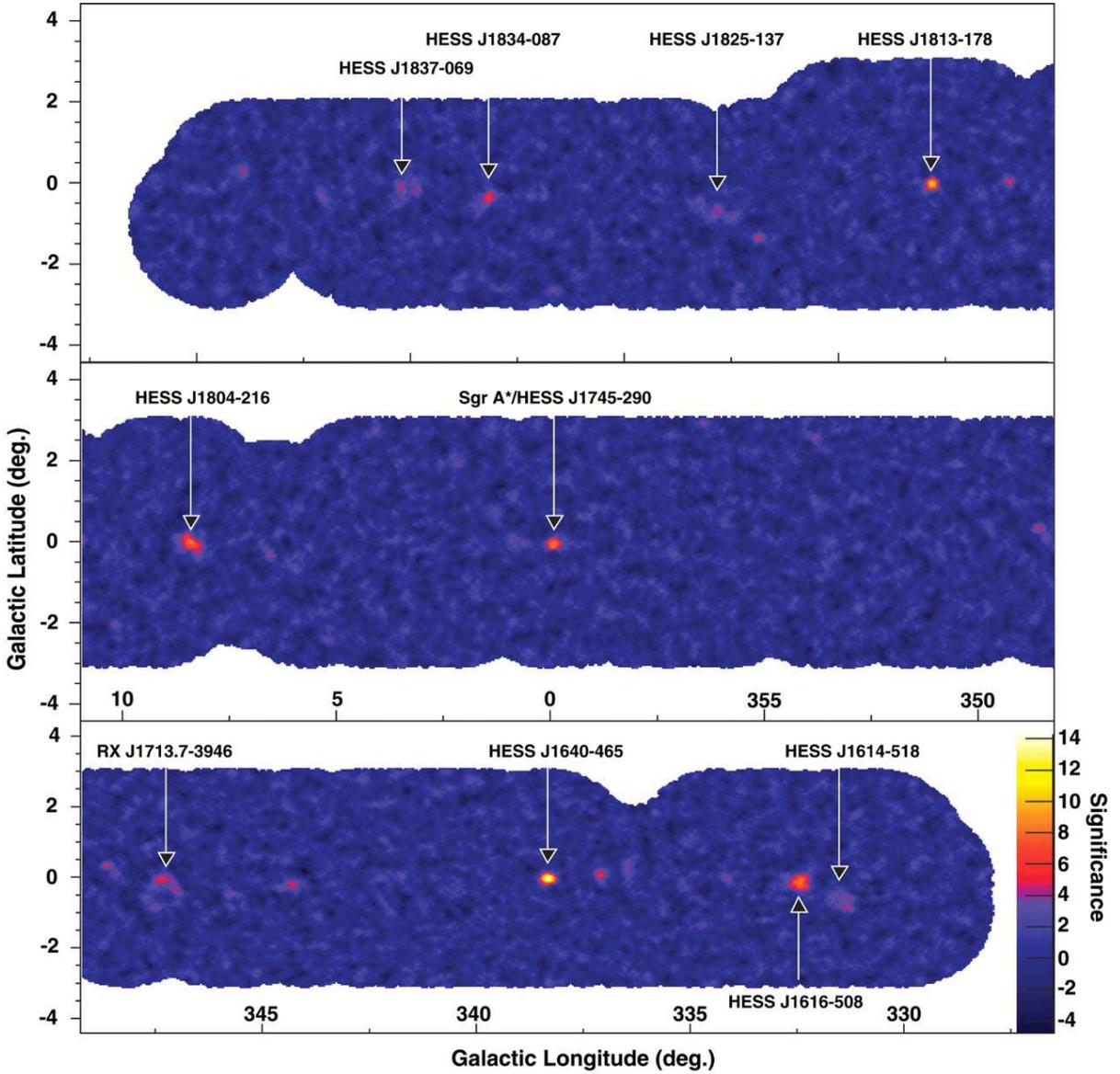}
\end{center}
\caption{\footnotesize { Significance map of the H.E.S.S. 2004
Galactic plane scan.  Reobservations of candidates from the initial
scan are included here. The background level was estimated using a
ring around the test source position~\cite{HESS2155}. On-source counts
are summed over a circle of radius $\theta$, where $\theta =
0.14^{\circ}$, a cut appropriate for point-like sources.  To calculate
the significances, it is necessary to correct for the relative
acceptance and area of the on and off regions. The correction is given
by a run-wise radially symmetric acceptance function, which describes
the background at the 1\% level.  Events falling up to 2$^{\circ}$
from the pointing direction of the system are used.  At this angle the
$\gamma$-ray acceptance efficiency has decreased to 20\% of its peak
value. After acceptance correction the approach of Li and
Ma~\cite{LiMa} is used to calculate the significance at every point on
the map. We note that consistent results are obtained for these 8
sources using a completely independent calibration and analysis
chain~\cite{ModelMathieu}.}}
\vspace{-0.3cm}
\label{fig:ScanMap}
\end{figure}

Eight unknown VHE sources were detected at the level of $>6$ standard
deviations after accounting for all trial factors (
Fig.~\ref{fig:ScanMap} and Table~\ref{tab:Hotspots}). The known VHE
sources HESS\,J1745-290 (at the Galactic Centre) and RX\,J1713.7-3946
are also detected in the scan data. The analysis used the standard
H.E.S.S. analysis procedure, optimised for point-like sources
~\cite{HESS2155}.

The significance values shown in Fig.~\ref{fig:ScanMap} (and $S_{3}$
of Table~\ref{tab:Hotspots}) do not directly reflect the probability
that a given signal represents a $\gamma$-ray source, because the
large number of search points in the sky map (250000) enhance the
chance for statistical fluctuations to mimic a signal.  Probabilities
must be scaled by a trial factor accounting for the number of
attempts to find a source, here conservatively assumed to be equal to
the number of points in the sky map. Simulations of randomly generated
sky maps without $\gamma$-ray sources show that the effective number
of trials is smaller than the number of points because of correlations
between adjacent search points, which are more closely spaced than the
instrumental width of the point spread function.  Trial factors apply
only for the initial search, where source candidates were identified
($S_{1}$, before trials), but not for follow-up observations, where
the significance ($S_{2}$) of a signal was evaluated assuming the
position derived from the initial search. The scaled-down significance
from the initial search and the result of the follow-up observations
were combined by summation in quadrature to give a final detection
significance ($S_{4}$).

Because most sources appear moderately extended, $S_{5}$ in
Table~\ref{tab:Hotspots} lists the significance similar to $S_{4}$ but
assuming extended sources . For each of these candidates a spectral
analysis was performed and Table~\ref{tab:Hotspots} also gives the
best fit flux above $200$\,GeV.

For the purpose of source size and position fitting, an additional cut
requiring an image size exceeding 200 photoelectrons in each camera
was applied. This cut further supresses the cosmic-ray background (by
a factor of $\sim 7$), at the expense of reduced statistical
significance and increased energy threshold (250~GeV). It also
improves the angular resolution by 30\%. After applying this cut the
spatial distribution of excess events for each candidate was fit to a
model of a two-dimensional Gaussian $\gamma$-ray emission region
(Table~\ref{tab:Hotspots}) convolved with the point spread function of
the instrument (derived from Monte Carlo simulations and verified by
observations of the Crab Nebula).

\begin{table}[h]
\begin{center}
\begin{tabular}{|l|c|c|c|c|c|c|c|c|c|} \hline 
Name& \multicolumn{2}{|c|}{Position} & Size $\sigma$ &
\multicolumn{5}{|c|}{Significance} & Flux  \\
& l & b & (arcmin.)  & $S_{1}$& $S_{2}$& $S_{3}$ & $S_{4}$ &
$S_{5}$ & \\ \hline\hline
HESS J1614-518$^\dag$ & $331.54^\circ$ & $-0.59^\circ$ & 12 & 5.2  & 4.3 & 6.7 & 4.7 &  6.8& $9$ \\ 
HESS J1616-508 & $332.40^\circ$ & $-0.15^\circ$ & 11 & 7.4  & 8.9 & 11.6 & 10.5 & 12.8 &$17$  \\
HESS J1640-465 & $338.32^\circ$ & $-0.02^\circ$ & 2 & 11.7 & 8.3 & 14.3 & 13.4 & 11.5 &$19$  \\
HESS J1804-216 &  $8.44^\circ$  & $-0.05^\circ$ & 13 & 8.2 &  5.9 & 10.1 & 8.8 & 9.6& $16$ \\ 
HESS J1813-178 & $12.81^\circ$  & $-0.03^\circ$ & 3  & 10.2  & 8.3 & 13.2 & 12.2 & 8.9 & $12$ \\
HESS J1825-137$^\dag$ & $17.78^\circ$  & $-0.72^\circ$ & 10 & 4.4 & 3.7 & 5.8 & 3.7 & 6.5 &  $9$ \\
HESS J1834-087 & $23.28^\circ$  & $-0.34^\circ$ & 12 & 6.7 & 5.6 & 8.7 & 7.2 & 7.8 & $13$ \\ 
HESS J1837-069 & $25.21^\circ$  & $-0.12^\circ$ & 4 & 6.0 & 6.0 & 8.4 & 6.9 & 6.4& $9$  \\
\hline
\end{tabular}
\vspace{-0.5cm}
\end{center}
\caption{\footnotesize{ Characteristics of the new $\gamma$-ray
  sources. Position: Galactic Longitude (l) and Latitude (b) with a
  statistical error in the range of 1-2 arcmin. Size: estimated source
  extension $\sigma$ for a brightness distribution of the form $\rho
  \propto exp(-r^{2}/2\,\sigma^{2})$ with a statistical error in the
  range of 10-30\%.  $S_{1}$: Significance for a point source cut of
  $\theta^2 = (0.14^{\circ})^2$, using scan data only, without
  correction for the number of trials. $S_{2}$: As for $S_{1}$ but
  only including follow-up observations of this source (no correction
  needed).  $S_{3}$: Significance of combined scan and follow-up
  observations (as shown in Fig.~\ref{fig:ScanMap}). $S_{4}$: As
  $S_{3}$ but including a correction for the number of trials ($n =
  250000$). $S_{5}$: As for $S_{4}$ but with an extended cut of
  $\theta^2 = (0.22^{\circ})^2$. Flux: Estimated flux above 200\,GeV
  $(\times 10^{-12}$cm$^{-2}$s$^{-1}$) with a statistical error
  between 10-35\%. $^\dag$: These sources were re-observed within the
  field of view of dedicated observations of another target.}}
\label{tab:Hotspots}
\end{table}

\begin{figure}
\begin{center}
\includegraphics[width=0.9\textwidth]{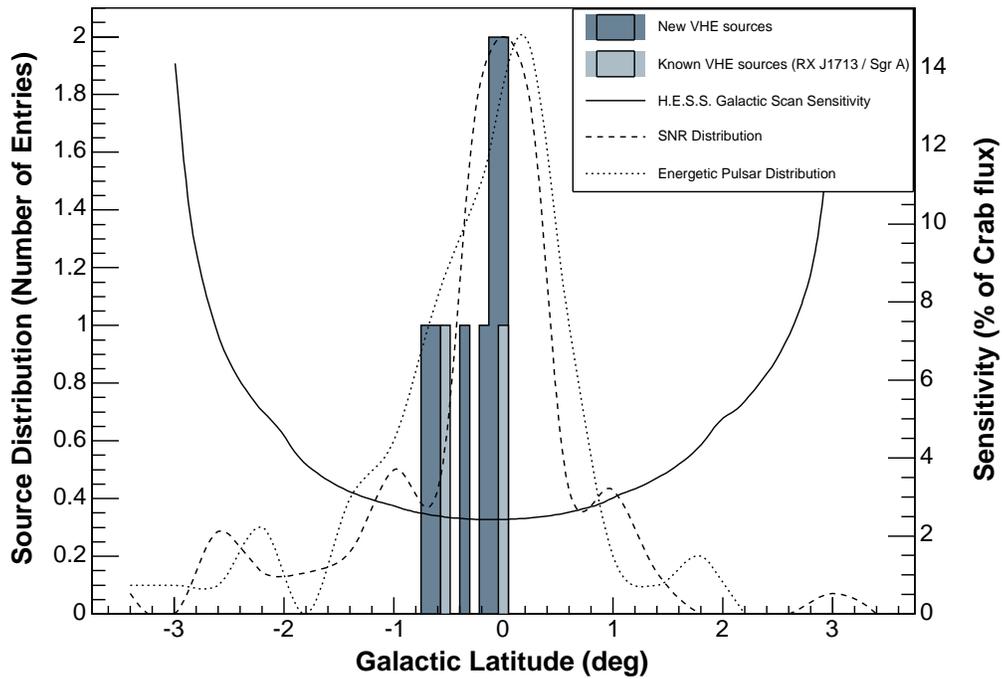}
\end{center}
\caption{\footnotesize Latitude distribution of the eight new VHE
  $\gamma$-ray sources (and two known sources in the scan region),
  along with the average sensitivity of the H.E.S.S. Galactic plane
  scan (for a 5 $\sigma$ detection, expressed as a percentage of the
  flux from the Crab Nebula). The distribution of Galactic SNRs and of
  energetic pulsars (including only pulsars with spin-down luminosity
  $\dot{E}$ more than $10^{34}$ erg/s.) are shown for comparison (for
  both distributions only objects within the longitude range of the
  HESS survey (-30$^\circ$$<$l$<$30$^\circ$ ) were selected).}
\label{fig:Latitude}
\end{figure}

The new Galactic VHE $\gamma$-ray emitters cluster close to the
Galactic plane (with a mean $b$ of -0.25$^\circ$ and a root mean
square (rms) of 0.25$^\circ$) (Fig.~\ref{fig:Latitude}). This is a
clear indication that we see a population of Galactic (rather than
extragalactic) sources.  Furthermore, the observed distribution
resembles that of Galactic SNRs~\cite{Greens} and of energetic pulsars
($\dot{E} > 10^{34}$ erg/s)~\cite{ATNF}.  However because of the
nonuniform exposure of the scan, and the unknown luminosity
distribution of the parent population, we cannot make a quantitative
statement on the compatibility of these distributions.

All of the sources are significantly extended beyond the size of the
H.E.S.S. point spread function. Given that the search described here
was optimised for point-like sources, it is likely that several more
significant sources with an extended nature exist in this dataset. We
note that the new sources mostly have spectra with rather hard photon
indices in the range between that expected for SNRs and that of the
Crab Nebula.

\begin{figure}
\begin{center}
\includegraphics[width=\textwidth]{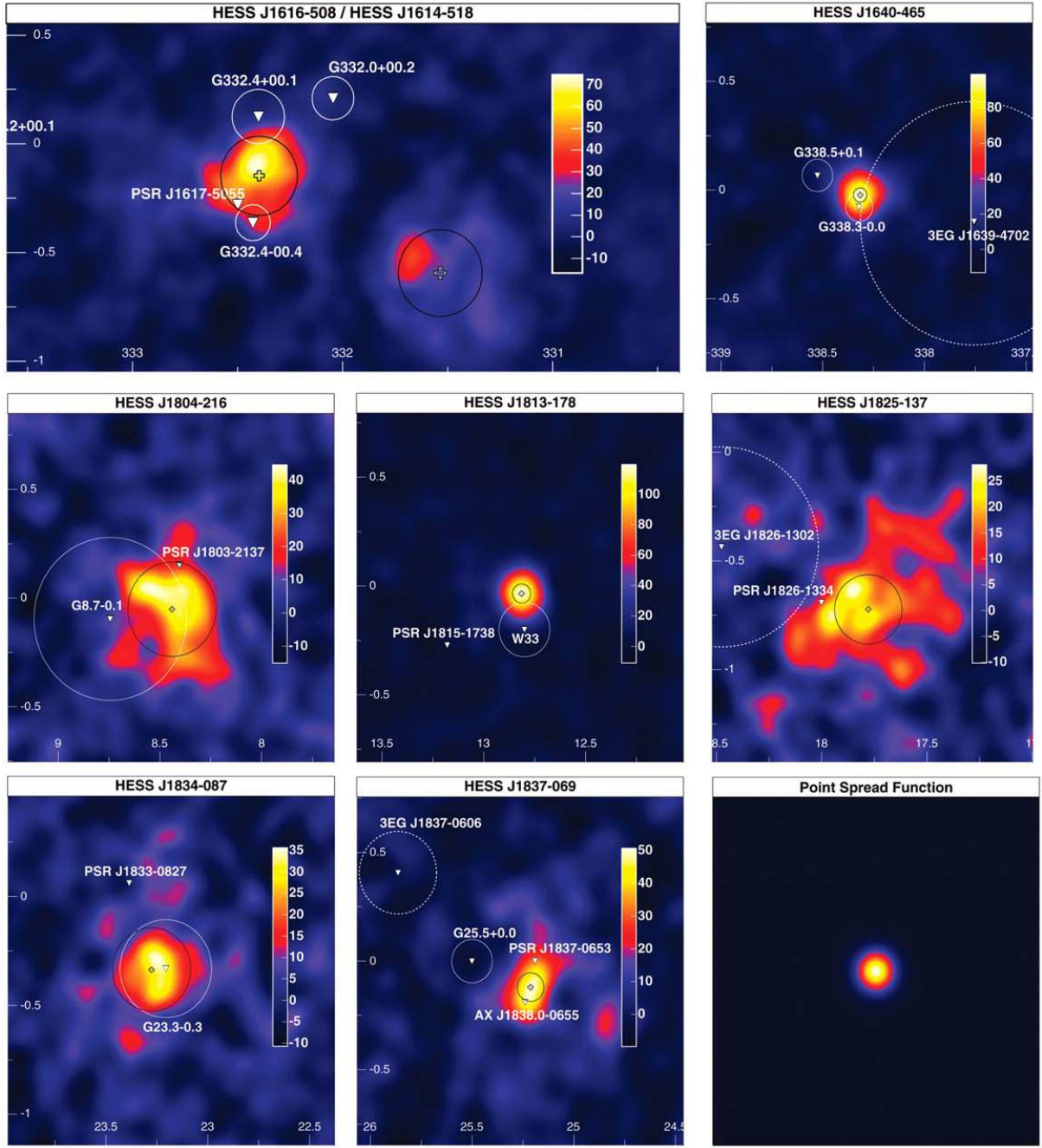}
\end{center}
\caption{\footnotesize Smoothed excess maps in units of counts of the
  regions around each of the eight new sources in Galactic coordinates
  (in degrees). An image size cut ($> 200$ photoelectrons) has been
  applied to reduce the background level and improve the angular
  resolution. A Gaussian of rms 0.05$^\circ$ is used for smoothing to
  reduce the impact of statistical fluctuations.  The best-fit
  centroids for the $\gamma$-ray emission are shown as crosses, and
  the best-fit rms size as a black circle.  Possible counterparts are
  marked by white triangles, with circles indicating the nominal
  source radius (or the position error in the case of EGRET sources).
  The lower right panel indicates the simulated point spread function
  of the instrument for these data, smoothed in the same way as the
  other panels.  }
\label{fig:Hotspots}
\end{figure}

We have searched for counterparts for the new $\gamma$-ray sources in
other wavelength bands. Fig.~\ref{fig:Hotspots} shows the $\gamma$-ray
emission from the region around each source together with potential
counterparts. Although the chance probability for spatial coincidences
with SNR in the region of the scan is not negligible (6\%), plausible
associations exist for three of the new sources -- HESS\,J1640-465,
HESS\,J1834-087 and HESS\,J1804-216 -- with an SNR. HESS\,J1640-465 is
spatially coincident with the SNR G338.3-0.0, which is a broken-shell
SNR lying on the edge of a bright ultra-compact HII
region~\cite{Greens}. This HII region could conceivably provide target
material for nuclear particles accelerated in the SNR, generating VHE
$\gamma$-rays by $\pi^{0}$ decay. The unidentified EGRET source
3EG\,J1639-4702 ~\cite{EGRET} lies 35 arc min away and could also be
connected to HESS\,J1640-465, given the position uncertainty of 34 arc
min of the EGRET source. HESS\,J1834-087 is spatially coincident with
SNR G23.3-0.3 (W\,41), a shell-type SNR of 27 arc min diameter
~\cite{W41Radio}. HESS\,J1804-216 coincides with the southwestern rim
of the shell-type SNR G8.7-0.1 (W30) of radius 26 arc min. From CO
observations~\cite{G8.7CO}, the surrounding region is known to be
associated with molecular gas where massive star formation is taking
place. This SNR could be associated with the nearby (25 arc min),
young pulsar PSR\,J1803-2137~\cite{PSRJ1803Xrays}.

HESS\,J1804-216 is one of three plausible associations with nebulae
powered by sufficiently energetic young pulsars. The others are
HESS\,J1825-137 and HESS\,J1616-508. HESS\,J1825-137 lies 13 arc min
from the pulsar PSR\,J1826-1334 which has been associated with the
close-by unidentified EGRET source
3EG\,J1826-1302~\cite{PSRJ1826-1334Egret}. A diffuse emission region,
5 arc min in diameter and extending asymmetrically to the south of the
pulsar was detected using the XMM (X-ray Multi-Mirror misson) x-ray
telescope~\cite{PSRJ1826-1334XMM}. This diffuse emission is
interpreted as synchrotron emission produced by a PWN. The VHE
emission is similarly located south of PSR\,J1826-1334, and could be
coincident with the PWN. The conversion efficiency implied from spin
down power to VHE $\gamma$-rays is less than 1\%. HESS\,J1616-508 is
located in the middle of a complicated region 9 arc min from the young
hard X-ray pulsar PSR\,J1617-5055~\cite{PSRJ1617Detection}.  The VHE
$\gamma$-ray flux is again less than 1\% of the spin-down luminosity
of the pulsar. The SNR G332.4-0.4 (RCW\,103)~\cite{Greens}, which
harbours a compact X-ray source
(1E\,161348-5055)~\cite{1E1613Discussion} lies 13 arc min away, as
does the SNR G332.4+0.1 (Kes\,32)~\cite{G332.4+0.1Chandra}, but
neither of these SNRs is spatially coincident with the VHE emission.
(Noted that none of the discussed SNR or PWN associations currently
meet the criteria necessary for promotion to counterparts.
Counterpart identification requires positional agreement with an
identified source, a plausible $\gamma$-ray emission mechanism and
consistent multi-frequency behaviour).

For HESS\,J1837-069 a potential counterpart is the diffuse hard X-ray
source G25.5+0.0, which is 12 arc min across and was detected in the
Advanced Satellite for Cosmology and Astrophysics (ASCA) Galactic
plane survey~\cite{BAMBA}. The nature of this bright X-ray source is
unclear but it may be an X-ray synchrotron emission dominated SNR such
as SN\,1006 or a PWN.  The brightest feature in the ASCA map
(AX\,J1838.0-0655), located south of G25.5+0.0, coincides with the
centre of gravity of the VHE emission and is therefore the most
promising counterpart candidate. This still unidentified source was
also serendipitously detected by BeppoSAX x-ray satellite instrument
and also in the hard X-ray (20-100 keV) band in the Galactic plane
survey performed with the Integral (International Gamma-Ray
Astrophysics Laboratory) satellite~\cite{1838Integral}.

For the two remaining sources HESS\,J1813-178 and HESS\,J1614-518, no
plausible counterparts have been found at other wavelengths.
HESS\,J1813-178 is not spatially coincident with any counterparts in
the region but lies 10 arc min from the centre of the radio source
W\,33. W\,33 extends over 15 arc min with a compact radio core
(G12.8-0.2) that is 1 arc min across~\cite{W33Radio}. The region is
highly obscured and has indications of recent star
formation~\cite{UltracompactHII}.  In its extended emission and
location close to an association of hot O and B stars, HESS\,J1813-178
resembles the unidentified TeV source discovered by HEGRA,
TeV\,J2032+4130~\cite{HegraUnid} and the first H.E.S.S. unidentified
$\gamma$-ray source HESS\,J1303-63~\cite{HESS1303}. HESS\,J1614-518
has no plausible SNR or pulsar counterpart. This source is in the
field of view of HESS J1616-508 which is located nearby ($\sim
1^\circ$ away).  The lack of any counterparts for these two sources
suggests the exciting possibility of a new class of `dark' particle
accelerators in the Galaxy.

In conclusion, we have on the basis of the survey generated a first
unbiased catalogue of very high energy $\gamma$-ray sources in the
central region of our Galaxy, extending our multi-wavelength knowledge
of the Milky Way into the VHE domain.  Three of the eight newly
discovered sources are potentially associated with supernova remnants,
two with EGRET sources. In three cases an association with
pulsar-powered nebulae is not excluded. At least two sources have no
identified counterpart in radio or X-rays, which suggests the exciting
possibility of a new class of `dark' nucleonic particle accelerators.
This catalogue provides insights into particle acceleration in our
Galaxy and adds a piece to the long-standing puzzle of cosmic-ray
origin.

\bibliography{scibib}

\begin{thebibliography}{}

\bibitem{MilagroAllSky} R. Atkins et al., Astrophys. J. \textbf{608}, 680 (2004).
\bibitem{TibetAllSky} M. Amenomori et al., Astrophys. J. \textbf{580}, 887 (2002).
\bibitem{HEGRA} A. Daum et al., Astropart. Phys. \textbf{8}, 1 (1997).
\bibitem{HEGRASCAN} F. Aharonian et al., Astron. Astrophys. \textbf{395}, 803 (2002). 
\bibitem{HESS} W. Benbow et al., in International Symposium on
  High-Energy Gamma-Ray Astronomy, F.~A. Aharonian, H.~J. V\"olk,
  D. Horns, Eds (AIP Conference Proceedings 745, 2005), pp 611-616.
\bibitem{HESSOPT} K. Bernl\"ohr et al., Astropart. Phys. \textbf{20}, 111 (2003). 
\bibitem{HESSCamera} P. Vincent et al., in Proceedings of the 28th
  International Cosmic Ray Conference, T. Kajita et al.,
  Eds. (Universal Academy Press, Tokyo, 2003), pp. 2887-2890.
\bibitem{HESSTrigger} S. Funk et al., Astropart. Phys. \textbf{22}, 285
  (2004).
\bibitem{HESSDesign2} F. Aharonian et al., Astropart. Phys. \textbf{6}, 369 (1997).
\bibitem{HESSGalCen} F. Aharonian et al., Astron. Astrophys. \textbf{425}, L13 (2004).
\bibitem{HESS1713} F. Aharonian et al., Nature \textbf{432}, 75 (2004).
\bibitem{HESS2155} F. Aharonian et al., Astron. Astrophys.,
  \textbf{430}, 865 (2005).
\bibitem{Greens} D.~A. Green, Bull. Astron. Soc. India \textbf{32}, 335 (2004).
\bibitem{ATNF} R. N. Manchester, G.~B. Hobbs, A. Teoh, M. Hobbs,
  Astron. J., in press (available at http://xxx.lanl.gov/abs/astro-ph/0412641).
\bibitem{EGRET}  R.~C. Hartman, et al. Astrophys. J. Suppl. Ser. \textbf{123}, 79 , 1999.
\bibitem{W41Radio} N.~E. Kassim, Astron. J. \textbf{103}, 943 (1992). 
\bibitem{G8.7CO} L. Blitz, M. Fich, A.~A. Stark,
  Astrophys. J. Suppl. Ser. \textbf{49}, 183 (1982).
\bibitem{PSRJ1803Xrays} N.~E. Kassim, Nature \textbf{343}, 146 (1990).  
\bibitem{PSRJ1826-1334Egret}P.~L. Nolan, W.~F. Tompkins,
  I.~A. Grenier, P.~F. Michelson, Astrophys. J. \textbf{597}, 615
  (2003). 
\bibitem{PSRJ1826-1334XMM}B.~M. Gaensler, N.~S. Schulz, V.~M. Kaspi,
  M.~J. Pivovaroff, W.~E. Becker, Astrophys. J. \textbf{588}, 441
  (2003).
\bibitem{PSRJ1617Detection} K. Torii et al., Astrophys. J. \textbf{494}, L207 (1998).  
\bibitem{1E1613Discussion} E.~V. Gotthelf, R. Petre, U. Hwang,
  Astrophys. J. \textbf{487}, L175 (1997).
\bibitem{G332.4+0.1Chandra} J. Vink, Astrophys. J. \textbf{604}, 693 (2004). 
\bibitem{BAMBA} A. Bamba, M. Ueno, K. Koyama, S. Yamauchi,
  Astrophys. J. \textbf{589}, 253 (2003). 
\bibitem{1838Integral} A. Malizia et al., in Proceedings of the V
  INTEGRAL Workshop, Munich, 16 to 20 February 2004 (available at
  http://xxx.lanl.gov/abs/astro-ph/0404596). 
\bibitem{W33Radio} A. Haschick, P.~T.~P. Ho, Astrophys. J. \textbf{267}, 638 (1983).
\bibitem{UltracompactHII}E. Churchwell, Astron. Astrophys. Rev. \textbf{2}, 79 (1990).
\bibitem{HegraUnid} F. Aharonian et al., Astron. Astrophys. \textbf{393}, L37 (2002).  
\bibitem{HESS1303} M. Beilicke et al.,, in International Symposium on
  High-Energy Gamma-Ray Astronomy, F.~A. Aharonian, H.~J. V\"olk,
  D. Horns, Eds (AIP Conference Proceedings 745, 2005), pp 347-352.
\bibitem{LiMa} T. Li, Y. Ma, Astrophys. J. \textbf{272}, 317 (1983).
\bibitem{ModelMathieu} M. de Naurois et al., in Proceedings of the 28th
  International Cosmic Ray Conference, T. Kajita et al.,
  Eds. (Universal Academy Press, Tokyo, 2003), pp. 2907-2910.

\end{thebibliography}
\bibliographystyle{Science}


\begin{scilastnote}
\item The support of the Namibian authorities and of the University of Namibia
in facilitating the construction and operation of H.E.S.S. is gratefully
acknowledged, as is the support by the German Ministry for Education and
Research (BMBF), the Max Planck Society, the French Ministry for Research,
the CNRS-IN2P3 and the Astroparticle Interdisciplinary Programme of the
CNRS, the U.K. Particle Physics and Astronomy Research Council (PPARC),
the IPNP of the Charles University, the South African Department of
Science and Technology and National Research Foundation, and by the
University of Namibia. We appreciate the excellent work of the technical
support staff in Berlin, Durham, Hamburg, Heidelberg, Palaiseau, Paris,
Saclay, and in Namibia in the construction and operation of the
equipment.
\end{scilastnote}
\end{document}